\documentclass[copyright,creativecommons,noncommercial]{eptcs}

\usepackage{underscore}           % Only needed if you use pdflatex.

\usepackage{todonotes}
\usepackage{url}
\usepackage[inline]{enumitem}
\usepackage{wrapfig}
\usepackage{clistings}

\makeatletter \let\lmtt@use@light@as@normal\@empty \makeatother

\newcommand{\ie}{i.e.}
\newcommand{\eg}{e.g.}
\newcommand{\qc}{QuickCheck}
\newcommand{\qv}{Quviq}
\newcommand{\autosar}{Autosar}

\newcommand{\lstdispfont}{%\small
     \ttfamily}

\title{Modelling of \autosar\ Libraries for Large Scale Testing\thanks{%
This work is supported by the Swedish Knowledge Foundation grant for
the AUTO-CAAS project.}}
\author{Wojciech Mostowski
\institute{Centre for Research on Embedded Systems\\Halmstad University, Sweden}
\email{wojciech.mostowski@hh.se}
\and
Thomas Arts
\institute{\qv\ AB, Sweden}
\email{thomas.arts@quviq.com}
\and
John Hughes
\institute{Department of Computer Science and Engineering\\
Chalmers University of Technology, Sweden}
\institute{\qv\ AB, Sweden}
\email{rjmh@chalmers.se, john.hughes@quviq.com}
}

\def\titlerunning{Modelling of \autosar\ Libraries for Large Scale Testing}

\def\authorrunning{W. Mostowski, T. Arts, J. Hughes}

\begin{document}

\maketitle

\begin{abstract}
We demonstrate a specific method and technology for
model-based testing of large software projects with the \qc\
tool using property-based specifications. Our specifications are very
precise, state-full models of the software under test (SUT).
In our approach we define (a)~formal descriptions of valid function call
sequences (public API), (b)~postconditions that check
the validity of each call, and (c)~\emph{call-out} specifications that
define and validate external system interactions (SUT calling external
API).  The \qc\ tool automatically generates and
executes tests from these specifications.
Commercially, this method and
tool have been used to test large parts of the
industrially developed automotive libraries based on the \autosar\
standard. In this paper, we exemplify our approach with a circular
buffer specified by \autosar, to demonstrate the
capabilities of the model-based testing method of \qc. Our example is
small compared to the commercial \qc\ models, but faithfully addresses many of the
same challenges.
\end{abstract}

\section{Introduction}

The multi-vendor and multi-platform
character of automotive system development is a major challenge in
testing the underlying software due to the multiplicity of possible
software and hardware interactions and differences between various
versions of the referenced libraries. Testing such software off-line
brings the additional challenge of substituting low-level hardware
interfaces and APIs with software mock-ups.

The \autosar\ standard~\cite{Autosar} is an
automotive industry effort to support multi-vendor software
development. In particular, the standard specifies the basic software
library interfaces and behaviour that \autosar\ compliant
implementations should adhere to. Ideally, compliant \autosar\
components from different vendors should be interchangeable with
little additional effort necessary to validate a complete system based
on \autosar. However, in practice small differences in implementation
choices due to performance considerations, particular platform
limitations, or interpretation freedom for (purposely or
inadvertently) underspecified behaviour may lead to compliance issues.

To manage the complexity of testing and validation in this context, we employ
Model-Based Testing (MBT) techniques~\cite{Tretmans11} rather than the
classical unit testing approach.  Because of the number of
test cases necessary to cover the interactions and behavioural
differences, unit testing suffers from scalability issues.  In
comparison, the base for MBT testing are models of the system
and its components, based on which tests in an arbitrary quantity 
are generated on the fly. A carefully chosen test data
generation method ensures a sufficient coverage (quality) of the resulting
tests. New configurations and library versions of the system under test
are simply covered by adapting the models and regenerating the
tests for various configurations. Another important feature of MBT,
as opposed to \eg\ model checking, is that an actual running
library or complete product compiled from a low-level language
can be tested and validated, rather than its model (although
in our particular case the model itself is also analysable, see 
Sect.~\ref{fig:cluster}). The infinite test space of this validation
scenario obviously cannot be completely covered by MBT, however,
the random test data generation ensures arguably good coverage of
the code under test. The specification and the testing can be
done with or without the source code availability for inspection,
the latter being often the situation in our industrial projects.

Our MBT method of choice is property based testing
using the \qc\ tool developed by
\qv~\cite{Arts+2006,Hughes2007}. In \qc, a model is an
Erlang~\cite{CesariniThompson2009} functional program that follows a predefined
structure and API to form a behavioural description of a given
software component. The properties that can be checked with generated
tests are postconditions (test oracles) of single operations and
\emph{call-out} specifications. Call-out specifications are given in a
process algebra-like language~\cite{MockingPaper} and describe the calls to other
components that the operation is allowed or required to make.  The
rest of the model defines the \emph{symbolic execution} state of the
component under test and operation preconditions, both of which define
the valid execution traces for the tests to be generated. For the
software under test written in C, which is the case for the
\autosar\ components, \qc\ provides a flexible interface that
allows to abstract the C function calls in the model and run the
underlying C~implementation during the actual testing.
The general philosophy of \qc\ is to run several short tests
quickly to detect problems early, rather than creating large and complex
testing scenarios for which failures are extremely difficult to analyse. 
Should a larger test case need to be generated to exhibit a bug,
\qc\ offers a mechanism to \emph{shrink} the test data to a smaller, possibly 
minimal, set that leads to the same failure. Consequently, the 
effort of debugging is also reduced.

\qc\ has been used in large scale to specify and test \autosar\
basic software implementations \cite{ArtsHNS15}.
In this paper we exemplify the \qc\ 
method by presenting a small model for one of the \autosar\ 
components, a circular buffer for queuing arbitrary
data. The client component is a
message box system, that uses the circular buffer to specifically
store message pointers (Sect.~\ref{sec:casestudy}).  Using this simple
example we show how the multi-level component specifications are
used to test \autosar\ implementations, and we show how C function
mocking is employed in \qc\ to test the call-out
specifications (Sect.~\ref{sec:implmodels}). Furthermore, we show how
\emph{clustering} specifications can be used to
validate interactions between components within one functional set,
\ie, that components respect each other APIs
within a cluster (Sect.~\ref{sec:clusters}). The concrete
implementation that we test comes from the
open-source \autosar\ implementation by
ArcCore\footnote{\url{https://www.arccore.com}.}, the third partner in
the AUTO-CAAS project.

The mocking mechanism of \qc\ for testing the call-out
specifications is also used for checking the failure consequences in
the complete system. This is done by mocking parts of the system with
a behaviour specified by a model, into which we can incorporate known
non-compliant behaviours or simply inject faults. This is described in
Sect.~\ref{sec:faults}.  
Then, Sect.~\ref{sec:conclusions} concludes the paper.  The
complete \qc\ model for this case study can be found in
App.~\ref{app:model} and on-line at \url{http://ceres.hh.se/mediawiki/CirqBuff_and_MBox_QuickCheck_Models}.

\section{The Case Study}
\label{sec:casestudy}

The case study we focus on is a circular buffer implementation 
\lstinline|CirqBuff| \cite{Autosar}.
The buffer has three core API
functions, defined by the following C~prototypes:
\begin{itemize}
\item \lstinline|CirqBufferType *CirqBuffDynCreate(size_t size, size_t dataSize);|
  that creates a circular buffer of the given size which stores
  arbitrary data (byte sequences) of a given size;
\item \lstinline|int CirqBuffPush(CirqBufferType *cPtr, void *dataPtr);|
  that enqueues an element into the buffer, the element is passed through
  a pointer to the array containing the data to be enqueued. The result
  is 0 on success, otherwise 1; and
\item \lstinline|int CirqBuffPop(CirqBufferType *cPtr, void *dataPtr);|
  that dequeues an element\break from the buffer by placing it in an array,
  and indicating the operation status with the return value.
\end{itemize}
The \lstinline|CirqBufferType| is a C structure that keeps track of the
current queue contents through the current element count \lstinline|currCnt| and \lstinline|head|\slash \lstinline|tail| pointers
to the allocated queue storage buffer \lstinline|bufStart| \dots \lstinline|bufEnd|,
as well as the the static limits of the buffer -- maximum size \lstinline|maxCnt|, 
and the \lstinline|dataSize| field, see Fig.~\ref{fig:cdef} for the complete C~structure definition. 
The data is copied to and from the buffer using
one of the \lstinline|memcpy| routines available on the target
platform, this is one of the configuration options of this component
that \autosar\ has multiplicity of, and that in general complicate the testing 
effort. 

\begin{figure}
\begin{lstlisting}[aboveskip=0pt,belowskip=0pt,basicstyle=\lstdispfont]
typedef struct {
    int maxCnt;                   /* The max number of elements in the list */
    int currCnt;                  /* The current number of elements */
    size_t dataSize;              /* Size of the elements in the list */
    void *head; void *tail;       /* List head and tail */
    void *bufStart; void *bufEnd; /* Buffer start/stop */
} CirqBufferType;
\end{lstlisting}
\caption{The \texttt{CirqBuffType} C data type definition.}\label{fig:cdef}
\end{figure}

This circular buffer is used in a few places in the open source
\autosar\ implementation, \eg, for communication transmission buffers. 
For presentation, we picked the simplest code available,
a message box component that
uses the circular buffer to store message pointers. The
\lstinline|Mbox| API is the following:
\begin{lstlisting}[basicstyle=\lstdispfont]
    Arc_MBoxType* Arc_MBoxCreate(size_t size);
    sint32 Arc_MBoxPost(const Arc_MBoxType *mPtr, void *msg);
    sint32 Arc_MBoxFetch(const Arc_MBoxType *mPtr, void *msg);
\end{lstlisting}
The only field of the \lstinline|Arc_MBoxType| structure is
\lstinline|cirqPtr| that keeps the circular buffer reference.
The post and fetch functions are each a simple
delegation of calls to \lstinline|CirqBufferPush| and
\lstinline|CirqBufferPop|, respectively, with the same meaning of result values. 
\lstinline|Arc_MBoxCreate| creates a circular buffer with the data size fixed to
the pointer size \lstinline|sizeof(void *)|. Thus,
\lstinline|MBox| specialises \lstinline|CirqBuff| to store 
pointers.

\section{Specifying and Testing the Implementations with \qc}
\label{sec:implmodels}

Tens of software components for vehicle software 
are specified in \autosar\ in a modular way.
The task at hand is to modularly test both \lstinline|CirqBuff| 
(Sect.\ \ref{test:cirqbuff}) and
\lstinline|Mbox| (Sect.~\ref{sec:mbox}) implementations. However, we also want to test
the software together, \ie, the component integration. We do so by automatically
deriving its specification (Sect.~\ref{sec:clusters}).  

Using MBT, one first builds a model,
the specified behaviour of the software under test (SUT). Based
on the model, an arbitrary 
number of test cases can be generated. Test cases are sequences of
calls to the API functions with generated data parameters. In
addition, the test contains information of what external API calls the
SUT makes. These API calls are automatically mocked. 
The generated 
tests are then executed on the SUT and the tool
validates postconditions and call-out behaviour for each call.

As MBT tool we use \qv's \qc\ with two distinguishing features: 
expressive, state-full, \emph{executable} models, and failed test minimisation, 
called \emph{shrinking}. Shrinking is a mechanism for reducing the 
size of the test case that failed, so that it still leads to the failure,
but with a possibly (much) shorter execution trace.
The controllable (see Sect.~\ref{sec:faults}) \emph{data generators} make this possible.

\lstset{language=Erlang,deletekeywords={size},morekeywords={record,andalso,begin}}

\subsection{The \texttt{CirqBuff} Model}
\label{test:cirqbuff}

\begin{figure}[t]
\begin{lstlisting}[aboveskip=0pt,belowskip=0pt,basicstyle=\lstdispfont]
-module(qc_cb).

-record(cb_state, {ptr, size, data_size, elements}).
initial_state() -> #cb_state{ elements=[] }.

invariant(S) -> length(S#cb_state.elements) =< S#cb_state.size.
postcondition_common(_S, {call, _, create, _, _ }, _Res) -> true;
postcondition_common(S, Call, Res) -> eq(Res, return_value(S, Call)).

create(Size, DSize) -> cb:'CirqBuffDynCreate'(Size, DSize).
create_args(_S) -> [nat(), nat()].
create_pre(S, [Size, _DSize]) -> S#cb_state.ptr == undefined andalso Size > 0.
create_next(S, R, [Size, DSize]) ->
    S#cb_state{ptr=R, size=Size, data_size=DSize }.

push(Ptr, Val) -> DataPtr = eqc_c:create_array(unsigned_char, Val),
  CallRes = cb:'CirqBuffPush'(Ptr, DataPtr), eqc_c:free(DataPtr), CallRes.
push_args(S) -> [S#cb_state.ptr, vector(S#cb_state.data_size, char())].
push_pre(S) -> S#cb_state.ptr /= undefined andalso
  length(S#cb_state.elements) < S#cb_state.size.
push_return(_S, _Args) -> 0.
push_next(S, _R, [_Ptr, Val]) ->
    S#cb_state{elements = S#cb_state.elements ++ [Val]}.
\end{lstlisting}
\caption{Excerpt of the \lstinline|CirqBuff| model.}
\label{fig:cbmodel}
\end{figure}

A snapshot of the \qc\ model for \lstinline|CirqBuff| for discussing
is shown in Fig.~\ref{fig:cbmodel}, the complete model is quoted in
App.~\ref{app:model}. The model is a collection of
Erlang~\cite{CesariniThompson2009} 
functions describing the different aspects of the behaviour of the
component under test. First, the declared record type \lstinline|cb_state| 
holds the \emph{symbolic execution} state of the model, \ie, a
model view of the \lstinline|CirqBuff|'s current state.
It contains the pointer \lstinline|ptr| reference of
the freshly created buffer by the \lstinline|CirqBuffDynCreate|, the \lstinline|size|
and the \lstinline|data_size| requested during the creation, and the model contents
of the buffer \lstinline|elements|, an Erlang list of arbitrary
values. The state is initialised by
\lstinline|initial_state| which defines \lstinline|elements| to be an empty
list representing an empty buffer. The remaining fields of
\lstinline|cb_state| are set to \lstinline|undefined|. 

The \lstinline|invariant| function declares a condition to be checked 
throughout the execution of the \emph{actual} tests. Here it is defined somewhat
artificially, because its validity is guaranteed by the construction of the rest
of the model (it only refers to the model state, not the implementation state),
but in general it can
be used to check specific validity conditions of the implementation, \eg, 
the \lstinline|head| pointer to be within the bounds of the 
\lstinline|CirqBuff| storage buffer. Similarly,
the common postcondition is checked after every
\lstinline|CirqBuff| operation call during the actual test. Using Erlang
parameter matching, we state that no extra checks are necessary for the
\lstinline|create| call, the postcondition is \lstinline|true|, while all
the other calls should have their actual return values equal to the
ones specified by the model. These model return values are specified separately
for each operation, as we describe next.

For every API operation we declare a group of \emph{callback}
functions that define:
\begin{enumerate} %[label=(\alph*)]
\item How to call the C function to execute the actual test (the base callback);
\item How to generate its arguments (the \lstinline|_arg| callback);
\item What is the precondition, defined over the model state and
  operation arguments, for the operation to be eligible for execution (the \lstinline|_pre| callback);
\item What is its expected return value (the \lstinline|_return| callback),
and finally;
\item How to update the model state after the operation is executed (the \lstinline|_next| callback).
\end{enumerate}
Each operation can have its own specific postcondition defined with 
a \lstinline|_post| callback and a call-out callback
to specify internal and external API interactions, we describe this in the next
section. The obligatory callbacks are the base one and the
arguments one, through which \qc\ recognises that an operation should be part of
test generation. The others have default values if unspecified.

The \lstinline|create| arguments are two positive integers
indicating the queue size and data size. The queue size
is restricted to be strictly positive by the precondition of \lstinline|create|. Thus, all
data sets generated that have the size equal to 0 are rejected during test generation
process. The precondition also requires that \lstinline|create| is called only 
once per test when the queue is not yet initialised, \ie, when no queue pointer reference 
has been recorded in the model state. The new model state
after \lstinline|create| is one with the newly created pointer and the creation arguments 
recorded in the corresponding fields of the state record. 

The first argument of the \lstinline|push| operation is the pointer stored earlier in 
the model state. The second argument is an element to be enqueued,
here a random \lstinline|data_size| length sequence of
characters. The precondition limits the \lstinline|push| calls to initialised and not yet full
queues. Due to this precondition, the 
expected return value is 0. In the next state the 
newly enqueued element is recorded in the \lstinline|elements| list.
The \lstinline|pop| operation is a mirror of \lstinline|push| with no
new specification concepts, hence we skip it in Fig.~\ref{fig:cbmodel}.

The precondition, arguments, and the next-state callbacks are primarily used 
during test case generation, \ie, when the model is symbolically executed to
generate valid execution traces with sample input data. The base and the
return value callbacks, together with the postcondition and the invariant described
above, are used during the test execution to 
\begin{enumerate*}[label=(\alph*)]
\item trigger the actual C implementation call, and
\item check the correctness of the result. 
\end{enumerate*}
The model state is updated both during the test generation, and test execution for 
bookkeeping. In the former case we refer to it as \emph{symbolic state},
in the latter as \emph{dynamic state}.

The execution of \lstinline|create| involves a direct call to its
C implementation \lstinline|CirqBufDynCreate| with the generated arguments.
This wrapped C call is automatically generated by the \qc\ C interface
generation library \lstinline|eqc_c|. A particular instance is set up 
by calling:
\begin{lstlisting}[basicstyle=\lstdispfont]
    eqc_c:start(cb, [ {cppflags, "-std=c99"}, {c_src, "cirq_buffer.c"} ]).
\end{lstlisting}
However, the \lstinline|CirqBuffPush| C implementation expects
the data to be passed through a pointer, and this pointer has to be
created first from the pure data generated by \qc. This is done with
the \qc\ library function \lstinline|eqc_c:create_array| that delegates 
pointer creation to the C executable, which is then passed on 
to \lstinline|CirqBuffPush|, after which it can be immediately freed,
see again the specification of \lstinline|push| in Fig.~\ref{fig:cbmodel}.

\subsection{Testing \texttt{CirqBuff} with \qc}

\begin{figure}
% {r}{.4\textwidth}
%\vspace*{-\baselineskip}
\begin{lstlisting}[aboveskip=0pt,belowskip=0pt,basicstyle=\lstdispfont,mathescape]
prop_cb() -> ?FORALL(Cmds, commands(qc_cb),
   begin {_, _, Res} = run_commands(Cmds), Res == ok end).

> eqc:quickcheck(qc_cb:prop_cb()).
...$[\textsl{Repeated 100 times}]$...
OK, passed 100 tests
\end{lstlisting}
\caption{Testing \texttt{CirqBuff}.}
\label{fig:cbtest}
\end{figure}

The model-based test execution of a specified component in \qc\ is done through 
defining a test property and passing it on to \qc, see Fig.~\ref{fig:cbtest}.
A test sequence is generated with \lstinline|commands|
for the model specified in module 
\lstinline|qc_cb|. The generated test sequence is
executed with \lstinline|run_commands(Cmds)|. 
The repetitive generation of test cases is controlled with the 
\lstinline|?FORALL| construct, by default 100 random tests are generated. 
Additional parameters can be used in the property and the model, \eg,
to induce the creation of more or longer test cases or to collect specific test
statistic (\eg, requirement coverage \cite{AH-ICST2016}). In the 
simplest form we just state \lstinline|Res == ok| to check for the 
result of running each test. Passing this property to \qc\ results in
passing all 100 tests.

In fact, we can run thousands of tests that exercise the creation of
circular buffers with different sizes and data sizes. We push and pop
from these buffers without finding any error in the
implementation. Notably, with the specific model construction we do respect the 
interface and never try to overflow or underflow the buffer.

\subsection{The \texttt{MBox} Model}
\label{sec:mbox}

By essentially copying and pasting the specification, we could test \lstinline|MBox|
in exactly the same way. However, 
being a client implementation that uses \lstinline|CirqBuff| as a library 
we want to take a different approach that details this
dependency. We test  \lstinline|MBox| in isolation and use \emph{mocking}
for the \lstinline|CirqBuff| calls.

The general idea of mocking is to
execute a test of an upper level operation (here from \lstinline|Mbox|)
in an environment, where the lower level operations (here from \lstinline|CirqBuff|)
are replaced by \emph{emulated} (mocked) ones.
This enables tracing of the particular low-level calls
actually made (or omitted, in case something should be forbidden), and to check
the call arguments. It also facilitates testing 
of software that uses hardware interfaces that cannot be run outside of the target 
platform (\eg, ECU ports or A/D converters). 

Code mocking is generally a tedious task involving writing replacement code of the mocked functions.
\qc\ has
an interface that makes the process considerably easier and provides a flexible 
language with clearly defined semantics to specify the interactions with the
mocked component~\cite{MockingPaper}.  
Mocking of a component is enabled by providing its API description in the model,
such definition for \lstinline|CirqBufPush| is given in Fig.~\ref{fig:mboxmodel}.
With just the API, \qc\ replaces the \lstinline|CirqBuff| implementation 
with an emulated alternative.
The \lstinline|stored_type| and \lstinline|buffer| clauses indicate that the 
\lstinline|dataPtr| parameter should not be simplified from a pointer to just a value 
during mocking (which is the default behaviour), and because \lstinline[language=C]|void|
has no values, the type is changed to \lstinline[language=C]|unsigned char| to enable 
dereferencing values from pointers.

\begin{figure}
\begin{lstlisting}[basicstyle=\lstdispfont,mathescape]
-module(qc_mbox).

api_spec() -> #api_spec { language = c, mocking = eqc_mocking_c,
  modules  = [ #api_module{ name = mbox, functions = [
    #api_fun_c{ name = 'CirqBuffPush', ret = int,
      args = [ #api_arg_c{type = 'CirqBufferType *', name = cPtr, dir = in},
               #api_arg_c{ type = 'void *', name = dataPtr, dir = in,
                 stored_type = 'unsigned char *',
                 buffer = {true, "cPtr->dataSize"}} ] $\dots$ }.

post_args(S) -> [S#mbox_state.ptr, vector(?PTR_SIZE, char())].
post_callouts(_, [_, Value]) ->
    ?CALLOUT(mbox, 'CirqBuffPush', [?WILDCARD, Value], 0).
\end{lstlisting}
\caption{Excerpt of the \lstinline|MBox| model.}
\label{fig:mboxmodel}
\end{figure}

The specification of \lstinline|MBox| follows the same
pattern as in \lstinline|CirqBuff|, here we only underline the core differentiating
elements.  
The message enqueueing operation \lstinline|post| is specialised 
to store pointers, \ie, byte sequences of \lstinline|?PTR_SIZE| length exactly, thus data size is
not stored in the model state. The
\lstinline|_callouts| callback specifies the external calls of \lstinline|post|.
Here, a single call to \lstinline|CirqBufPush| is required with two arguments.
The first is the pointer to the \lstinline|CirqBuffType| structure that is created by 
\lstinline|MBox| earlier with a call to \lstinline|CirqBuff|'s create function. We choose to ignore it 
with \lstinline|?WILDCARD|, meaning that we do not check if \lstinline|MBox|
uses this pointer consistently with all \lstinline|CirqBuff| calls.
The second argument is the value pushed onto the circular buffer and it is simply passed on
from the call to \lstinline|post|.
The result of the call
should always be 0 indicating a success. 

Specified this way, \lstinline|MBox| can be tested 
with the \lstinline|CirqBuff| fully mocked and with all the calls to \lstinline|CirqBuff|
traced and checked. The \lstinline|MBox| test property is identical 
to the one for \lstinline|CirqBuff|, the difference in test generation and 
execution lies in the mocked API specified in the model. Should the calls to 
\lstinline|CirqBuff| made by \lstinline|MBox| be different from the specified ones, or missing,
the test fails and the details of the mismatch between the 
expected and actual calls is reported.

\subsection{Analysing Models and API Interactions through Clustering}
\label{sec:clusters}

We have specified only two components, one depending on the other, and
we have shown how the interactions down this simple hierarchy can be modelled and tested 
through mocking and call-out specifications. This scenario, however, is very simple.
Moreover, we only tested for the occurrence of the lower-level call in relation
to the state of the
higher-level model and the particular operation parameters. More generally, we would 
like to test multi-level dependencies of several components and check that they
all fully respect each other APIs according to the corresponding model 
specifications. For example, so far we did not checked 
that the \lstinline|MBox| is not trying to overflow the circular buffer
by placing too many elements in it.\footnote{Actually, incidentally it was checked 
implicitly, but only because of the specific model construction and the fact that the sizes 
of the two components are always the same. It was not done explicitly by checking the 
preconditions of the circular buffer.} 
A typical example of a scenario requiring such elaborate check 
is a protocol stack, in \autosar\ there are several such protocols specific to automotive
applications, including CAN-Bus and Flex-Ray.

To enable such a check, the set of interacting components is
placed in a so-called \emph{cluster} model. The models of
components placed in a cluster
only need slight extensions to facilitate the process. First,
the operations that can be called by other components have to name the possible
callers by their module name.
Second, the API specifications of the mocked calls to the lower-level routines
need to have a direct indication specified as to which modelled component and operation in 
the cluster they refer to.

\begin{figure}
%[12]{r}{.4\textwidth}
%\vspace{-.5\baselineskip}
\begin{lstlisting}[aboveskip=0pt,belowskip=0pt,basicstyle=\lstdispfont,mathescape]
-module(qc_cluster).

components() -> [qc_cb, qc_mbox].

api_spec() -> eqc_cluster:api_spec(?MODULE).

property_cluster() -> ?FORALL(Cmds, commands(?MODULE),
    begin {_, _, Res} = run_commands(Cmds), Res == ok end).
\end{lstlisting}
\caption{The cluster model.}
\label{fig:cluster}
\end{figure}

In our case study, each operation in the model of the
\lstinline|CirqBuff| has to name the \lstinline|MBox| model as the caller of the operation. 
For the \lstinline|push| operation this is done with  
\lstinline|push_callers() -> [qc_mbox].| Then, the mocked API 
for \lstinline|CirqBuffPush|
is extended with a \lstinline|classify = {qc_cb, push}| clause
to indicate the binding
to the appropriate operation in the \lstinline|CirqBuff| model. These two simple extensions
repeated for every operation
allow us to now give a short and straightforward cluster specification for the two components 
shown in Fig~\ref{fig:cluster}. It simply lists the components in the cluster and 
defines their 
underlying collective API by collecting the APIs of all included components. Testing the specified cluster property by \qc\ establishes that 
\lstinline|MBox| respects the API of \lstinline|CirqBuff|. 
Namely, if any of the \lstinline|MBox| calls is eligible to be included in the generated test 
case for \lstinline|MBox|, then the corresponding call-out to \lstinline|CirqBuff| is checked 
to be eligible in the \lstinline|CirqBuff| model. This includes checking that the caller 
is an authorised one. This is done by pure \emph{symbolic} execution of both 
models, no  actual implementations are tested in this process, it is only the models and
their interactions that are checked. The actual implementations are tested one by one using 
their corresponding models like described above. 

\section{Faults, Deviations, and Consequence Testing}
\label{sec:faults}

The mocking mechanism discussed in Sect.~\ref{sec:mbox} also allows us 
to test implementation against \emph{mutations} of sub-components
to test for fault tolerance. This is part of on-going 
research~\cite{AUTOCAASWASA15,WASA16}, here we exemplify the idea and show the capabilities of \qc\
to guide test generation to trigger a hidden fault. 

\looseness=1
Assume a scenario where the \lstinline|CirqBuff| has either faults, or deliberate 
modifications dictated by, \eg, performance optimisations or target platform limitations. 
These changes may or may not be compliant to the specification given by the standard. When they are not 
we call them \emph{deviations}, rather than just faults. Done in good faith, they do not necessarily have 
to lead to problems or faults in the top-level behaviour. We want to model and test for this,
without going through the burden of modifying the implementation 
of the low-level component, or constructing several implementations for this purpose. Instead, 
we test using the mocked version of the possibly deviating component
and introduce the deviation in the mocking. 

Suppose the \lstinline|CirqBuff| silently accepts all buffer sizes when creation is requested, however, 
due to platform limitations only buffers of size 128 are created, under the assumption that the client 
code can check the available space each time a new elements is enqueued (which in fact 
our particular client \lstinline|MBox| \emph{does not do}). To include this behaviour in our model
it is sufficient to make one line change in the call-out specification of \lstinline|CirqBuffPush|
to return \lstinline|1| on reaching the ``silent'' limit of 128 elements:
\begin{lstlisting}[basicstyle=\lstdispfont]
    ?CALLOUT(mbox, 'CirqBuffPush', [?WILDCARD, Value],
        if length(S#mbox_state.elements) < 128 -> 0; true -> 1 end)
\end{lstlisting}

Regenerating the tests with this updated model \emph{does not} reveal any problems in 
the implementation of \lstinline|MBox|, while it \emph{should}. The reason is the  
\lstinline|nat()| size generator that does not attempt to create \lstinline|MBox|-es of
sizes large enough, and even if it would, the generated traces do not 
contain enough \lstinline|post| operations to fill up the buffer and trigger the fault.
To stimulate \qc\ towards reaching the 
faulty state we first change the size data generator from \lstinline|nat()| to 
\lstinline|choose(1, 256)| that uniformly generates integers in the prescribed range rather than 
conservatively small ones. Then, we alter the test property with \lstinline|more_commands| to increase 
the expected length of generated test traces by the factor of 50,
and we provide relative \emph{weights} to ensure that the \lstinline|post| operation dominates in the generated tests to quickly fill up the buffer, see Fig.~\ref{fig:moretests}.
This is sufficient to discover the fault after a handful of test runs by \qc.
Expectedly, the shrinking algorithm is not able to generate test cases shorter than 
130 commands to
trigger this fault, however, 130 \emph{is} the minimal fault triggering trace
-- one \lstinline|create|, 128 \lstinline|post|-s to fill up the buffer, and one last to overflow it.

\begin{figure}
%{r}{.45\textwidth}
%\vspace{-\baselineskip}
\begin{lstlisting}[aboveskip=0pt,basicstyle=\lstdispfont,mathescape,deletekeywords={after}]
weight(#mbox_state{ptr=undefined}, Op) ->
    case Op of create -> 1; _ -> 0 end;
weight(_S, create) -> 0;
weight(_S, post) -> 5;
weight(_S, fetch) -> 1.

prop_mbox() -> ?FORALL(Cmds, more_commands(50, commands(?MODULE)), $\dots$

> eqc:quickcheck(qc_mbox:prop_mbox()).
...$\textsl{[snip]}$.Failed! After 23 tests.
$\textsl{[Long trace of 218 commands]}$
Shrinking xx$\textsl{[snip]}$.x(1056 times)
$\textsl{[Shrunk failure trace of 130 commands]}$ 
Reason:
  Post-condition failed:
  Failed postcondition: common: 1 /= 0
false
\end{lstlisting}
\caption{Extending the test scenario.}
\label{fig:moretests}
\end{figure}

Similarly to \lstinline|more_commands|, with the \lstinline|more_bugs| directive
\qc\ attempts to discover more than one fault in one testing sessions, instead of 
halting after the first test failure. This way, different kinds of faults
(different postconditions failing) can be identified sooner. 
 
An often quoted criticism towards MBT is the high specification to implementation 
ratio, here roughly 1:1 which is equivalent to a reimplementation effort. However, this is \emph{all} the 
effort required -- with one or two line changes to the model we have just shown how to test 
the implementation against a different sub-component and how to 
substantially change the shape of 
the generated test cases to trigger a hidden fault.
In the classical testing approach, a large number of new 
test cases would have to be developed as well as complete component implementations to achieve the
same effect. 

Finally, some remarks about the test execution and bug finding time performance should be made. 
The complete process of generating the basic 100 test cases for the \lstinline|MBox| implementation in its initial 
configuration from Sect.~\ref{sec:mbox} (\ie, with \lstinline|CirqBuff| mocking, but without  
custom operation weights and 
without \lstinline|more_commands|) and executing them takes a mere 3~seconds on a standard laptop. 
Expanding the 
test scenario to hit the bug increases this time to 36 seconds, which is also acceptable.
Moreover, this is the time for generating and executing all 100 tests, 
but the bug is found after just $\sim$20 tests in 7 seconds. Specific test 
scenarios that attempt to find an infrequently occurring bug may require increasing the standard 100 random tests towards a much larger number
(tens or hundreds of thousands). However, the running time will grow only linearly to the number of 
tests, which is perfectly acceptable considering the goal of finding the bug and the full process automation.

What does take considerable time is the shrinking process for large counterexamples. 
In the example we have shown in Fig.~\ref{fig:moretests} it already took just under 6 minutes for
the counterexample trace of 218 operations.
This is because each single shrinking attempt requires at least one complete rerun of one test case
to check if the bug still occurs with the smaller trace.
This procedure, however, is again 
fully automatic and replaces laborious, time consuming, manual process of analysing and rearranging test 
cases by the test engineer to find a shorter counterexample.

\section{Conclusions}
\label{sec:conclusions}

Using a small example from the automotive \autosar\
standard we have presented the \qc\ meth\-od\-ol\-o\-gy
for model-based testing. By generating tests from models, instead of
manually writing them, we can cover an arbitrary set of
test scenarios. By controlling test data generators, we can drive
the process towards a specific testing goal. The randomness
of test generation brings the additional advantage of creating
a-typical test sequences that would not be considered by a testing
engineer.

The added value of model-based testing is that
we can arbitrarily raise the number of tests to cover new
behaviours of the system under test with very little additional
effort.  In our small case study, changing a couple of lines in the
model exhibited a hidden bug, while the shrinking mechanism of
\qc\ provided a minimal counter example for the bug without extra
development cost, in favour of additional running time to crunch the
test data.

An indispensable feature of \qc\ is the flexible API mocking
mechanism.  Apart from enabling software abstractions of hardware
interfaces, which is crucial for off-line testing of embedded
software, we have demonstrated how it can be used to test for fault
tolerance of the complete system. We achieved this by injecting faults
in the component mock-ups, against which the system is tested. This
allows to test against many different (faulty or correct) component
implementations, that in fact do not have to and may not even exist,
provided a configurable model of the component is available.

\looseness=-1
Similar to writing test cases, building the models requires domain
knowledge and the ability to interpret the specifications written in
English. But by domain experts, it is conceived harder to write a
model than to write a single test case. This is not \qc\
specific, but a general observation that reasoning about all cases
at once is harder than reasoning about one specific case. When
the hurdle of writing models is overcome, a
second challenge, a technical one, is to learn the modelling
language. It takes a course and some weeks of experience before
engineers feel comfortable with modelling C~code in Erlang.

In the presented case study that mimics the real testing
effort of the \autosar\ software, the specification to implementation
ratio is 1:1. However, as the code grows towards the complete
\autosar\ library, the size of the model does not grow as fast. For
the complete set of \autosar\ libraries that has been tested
commercially~\cite{ArtsHNS15} the models are only about 20-25\% the
size of the feature rich C~source code under test. Obviously, 
specifying real \autosar\ libraries on a larger scale brings
additional challenges and benefits compared to what we have described
in the paper.

On the benefits side, we can create test suites for really complex
library configurations using the same models. This is not so obvious in our
example, but an \autosar\ configuration file consists of
thousands of parameters. Inspecting and updating manually written test
cases for such a parameter space is really hard. When using models, we
can do this automatically.

Another benefit that is hard to see in the paper, but showed valuable
in production is the clustering. A typical test case for the CAN bus
would cover five different specifications, from driver up to CAN State Manager
(\texttt{CanSM)} and CAN Interface (\texttt{CanIf}) \cite{Autosar-ATS}.
Each such specification is a
separate document. There is no detailed specification on
how the complete stack should behave. Therefore, engineers manually
constructing those tests must have all possible interactions in their
head. And indeed, this is error prone. 
We encoded the six test cases provided in \cite{Autosar-ATS} in our format, and
executed them against the model. We found that three of the six tests agreed with our
model (\ie, these tests would possibly be generated by our model);
the other three were not accepted by it. So we could never have
generated these tests -- but this turned out to be a \emph{good} thing:
the three standardised test cases failed to consider 
a specific transmission recovery scenario. 
This has been acknowledged as an error, and is filed in \autosar's Bugzilla
for correction in the next release. In other words, being able to compute
the complex interactions instead of trying to imagine them, is an
advantage.

On the challenges side when dealing with larger models is to get those
models generate valid test cases. There we used implementations from
different vendors to validate against and to discuss cases in which
our models and C implementation did not agree. In this process, 
ambiguities or underspecifications were found in the standard that lead to
differently behaving stacks, whereas they were all arguably correct. This
is a strength and weakness of low-level modelling; on the one hand one
wants to make all differences on the API level visible, on the other hand,
some design decisions that make an implementation faster or less
memory intensive, may be tolerated. The \qc\ methodology had problems in
tolerating ``that is also fine'' approach.

Another effort not visible in small, simple examples is finding good
values for test distribution. In the larger models we needed to add
certain distributions to get into some interesting states more often.
For example, if a large message is sent in chunks over the CAN bus,
one wants at least one test that finishes the complete transfer. If
the API calls are chosen \emph{just} at random,
this is unlikely to happen. Therefore,
such behaviour needs to be guided by test distribution values. 
 
% \bibliographystyle{eptcs}
% \bibliography{references}

\begin{thebibliography}{10}
\providecommand{\bibitemdeclare}[2]{}
\providecommand{\surnamestart}{}
\providecommand{\surnameend}{}
\providecommand{\urlprefix}{Available at }
\providecommand{\url}[1]{\texttt{#1}}
\providecommand{\href}[2]{\texttt{#2}}
\providecommand{\urlalt}[2]{\href{#1}{#2}}
\providecommand{\doi}[1]{doi:\urlalt{http://dx.doi.org/#1}{#1}}
\providecommand{\bibinfo}[2]{#2}

\bibitemdeclare{inproceedings}{AH-ICST2016}
\bibitem{AH-ICST2016}
\bibinfo{author}{T.~\surnamestart Arts\surnameend} \&
  \bibinfo{author}{J.~\surnamestart Hughes\surnameend} (\bibinfo{year}{2016}):
  \emph{\bibinfo{title}{How Well are Your Requirements Tested?}}
\newblock In: {\sl \bibinfo{booktitle}{2016 {IEEE} International Conference on
  Software Testing, Verification and Validation}}, pp.
  \bibinfo{pages}{244--254}, \doi{10.1109/ICST.2016.23}.

\bibitemdeclare{inproceedings}{Arts+2006}
\bibitem{Arts+2006}
\bibinfo{author}{T.~\surnamestart Arts\surnameend},
  \bibinfo{author}{J.~\surnamestart Hughes\surnameend},
  \bibinfo{author}{J.~\surnamestart Johansson\surnameend} \&
  \bibinfo{author}{U.~\surnamestart Wiger\surnameend} (\bibinfo{year}{2006}):
  \emph{\bibinfo{title}{Testing telecoms software with {QuviQ} {QuickCheck}}}.
\newblock In: {\sl \bibinfo{booktitle}{Proceedings of ERLANG'06}},
  \bibinfo{publisher}{ACM}, pp. \bibinfo{pages}{2--10},
  \doi{10.1145/1159789.1159792}.

\bibitemdeclare{inproceedings}{ArtsHNS15}
\bibitem{ArtsHNS15}
\bibinfo{author}{T.~\surnamestart Arts\surnameend},
  \bibinfo{author}{J.~\surnamestart Hughes\surnameend},
  \bibinfo{author}{U.~\surnamestart Norell\surnameend} \&
  \bibinfo{author}{H.~\surnamestart Svensson\surnameend}
  (\bibinfo{year}{2015}): \emph{\bibinfo{title}{Testing {AUTOSAR} software with
  QuickCheck}}.
\newblock In: {\sl \bibinfo{booktitle}{Eighth {IEEE} International Conference
  on Software Testing, Verification and Validation Workshops}}, pp.
  \bibinfo{pages}{1--4}, \doi{10.1109/ICSTW.2015.7107466}.

\bibitemdeclare{inproceedings}{AUTOCAASWASA15}
\bibitem{AUTOCAASWASA15}
\bibinfo{author}{T.~\surnamestart Arts\surnameend} \& \bibinfo{author}{M.R.
  \surnamestart Mousavi\surnameend} (\bibinfo{year}{2015}):
  \emph{\bibinfo{title}{Automatic Consequence Analysis of Automotive Standards
  ({AUTO-CAAS})}}.
\newblock In: {\sl \bibinfo{booktitle}{First International Workshop on
  Automotive Software Architectures ({WASA} 2015)}}, \bibinfo{publisher}{ACM
  Press}, pp. \bibinfo{pages}{35--38}, \doi{10.1145/2752489.2752495}.

\bibitemdeclare{misc}{Autosar}
\bibitem{Autosar}
\bibinfo{author}{\surnamestart {AUTOSAR Consortium}\surnameend}
  (\bibinfo{year}{2013}): \emph{\bibinfo{title}{AUTomotive Open System
  ARchitecture, standard documents}}.
\newblock \bibinfo{howpublished}{\url{https://autosar.org/}}.

\bibitemdeclare{misc}{Autosar-ATS}
\bibitem{Autosar-ATS}
\bibinfo{author}{\surnamestart {AUTOSAR Consortium}\surnameend}
  (\bibinfo{year}{2014}): \emph{\bibinfo{title}{{Acceptance Test Specification
  of Communication on CAN bus -- Release 1.0.0}}}.

\bibitemdeclare{book}{CesariniThompson2009}
\bibitem{CesariniThompson2009}
\bibinfo{author}{F.~\surnamestart Cesarini\surnameend} \&
  \bibinfo{author}{S.~\surnamestart Thompson\surnameend}
  (\bibinfo{year}{2009}): \emph{\bibinfo{title}{Erlang Programming}}.
\newblock \bibinfo{publisher}{O'Reilly}.

\bibitemdeclare{inproceedings}{Hughes2007}
\bibitem{Hughes2007}
\bibinfo{author}{J.~\surnamestart Hughes\surnameend} (\bibinfo{year}{2007}):
  \emph{\bibinfo{title}{{QuickCheck} testing for fun and profit}}.
\newblock In: {\sl \bibinfo{booktitle}{Proceedings of PADL'07}},
  \bibinfo{publisher}{Springer}, pp. \bibinfo{pages}{1--32},
  \doi{10.1007/978-3-540-69611-7_1}.

\bibitemdeclare{inproceedings}{WASA16}
\bibitem{WASA16}
\bibinfo{author}{S.~\surnamestart Kunze\surnameend},
  \bibinfo{author}{W.~\surnamestart Mostowski\surnameend},
  \bibinfo{author}{M.R. \surnamestart Mousavi\surnameend} \&
  \bibinfo{author}{M.~\surnamestart Varshosaz\surnameend}
  (\bibinfo{year}{2016}): \emph{\bibinfo{title}{Generation of Failure Models
  through Automata Learning}}.
\newblock In: {\sl \bibinfo{booktitle}{Second International Workshop on
  Automotive Software Architectures ({WASA} 2016)}}, \bibinfo{publisher}{IEEE
  Society}, pp. \bibinfo{pages}{22--25}, \doi{10.1109/WASA.2016.7}.

\bibitemdeclare{incollection}{MockingPaper}
\bibitem{MockingPaper}
\bibinfo{author}{J.~\surnamestart Svenningsson\surnameend},
  \bibinfo{author}{H.~\surnamestart Svensson\surnameend},
  \bibinfo{author}{N.~\surnamestart Smallbone\surnameend},
  \bibinfo{author}{T.~\surnamestart Arts\surnameend},
  \bibinfo{author}{U.~\surnamestart Norell\surnameend} \&
  \bibinfo{author}{J.~\surnamestart Hughes\surnameend} (\bibinfo{year}{2014}):
  \emph{\bibinfo{title}{An Expressive Semantics of Mocking}}.
\newblock In: {\sl \bibinfo{booktitle}{Fundamental Approaches to Software
  Engineering}}, {\sl \bibinfo{series}{LNCS}} \bibinfo{volume}{8411},
  \bibinfo{publisher}{Springer}, pp. \bibinfo{pages}{385--399},
  \doi{10.1007/978-3-642-54804-8_27}.

\bibitemdeclare{inproceedings}{Tretmans11}
\bibitem{Tretmans11}
\bibinfo{author}{J.~\surnamestart Tretmans\surnameend} (\bibinfo{year}{2011}):
  \emph{\bibinfo{title}{Model-Based Testing and Some Steps towards Test-Based
  Modelling}}.
\newblock In: {\sl \bibinfo{booktitle}{Formal Methods for Eternal Networked
  Software Systems}}, {\sl \bibinfo{series}{LNCS}} \bibinfo{volume}{6659},
  \bibinfo{publisher}{Springer}, pp. \bibinfo{pages}{297--326},
  \doi{10.1007/978-3-642-21455-4_9}.

\end{thebibliography}

\appendix
\section{The Complete \texttt{CirqBuff} and \texttt{MBox} QuickCheck Model in Erlang}
\label{app:model}

Below we present the complete model discussed in this paper. The snippets used for presentation
are slightly abbreviated and simplified for clarity, here the specifications are given in their
full form, also to be found at
\url{http://ceres.hh.se/mediawiki/CirqBuff_and_MBox_QuickCheck_Models}.

\lstset{belowskip=\bigskipamount,basicstyle=\small\ttfamily,%
  morekeywords={define,include,compile,include_lib}}

\subsection{File \texttt{source_path.hrl}} \lstinputlisting{./anc/source\string_path.hrl}

\subsection{File \texttt{qc_cb.hrl}} \lstinputlisting{./anc/qc\string_cb.hrl}

\subsection{File \texttt{qc_cb_setup.erl}} \lstinputlisting{./anc/qc\string_cb\string_setup.erl}

\subsection{File \texttt{qc_cb.erl}} \lstinputlisting{./anc/qc\string_cb.erl}

\subsection{File \texttt{qc_mbox.hrl}} \lstinputlisting{./anc/qc\string_mbox.hrl}

\subsection{File \texttt{qc_mbox_capi.erl}} \lstinputlisting{./anc/qc\string_mbox\string_capi.erl}

\subsection{File \texttt{qc_mbox_setup.erl}} \lstinputlisting{./anc/qc\string_mbox\string_setup.erl}

\subsection{File \texttt{qc_mbox.erl}} \lstinputlisting{./anc/qc\string_mbox.erl}

\subsection{File \texttt{qc_cluster.erl}} \lstinputlisting{./anc/qc\string_cluster.erl}

\end{document}